\documentclass[pre,twocolumn,showpacs,superscriptaddress] {revtex4}
\usepackage{graphicx}% Include figure files
\usepackage{dcolumn}% Align table columns on decimal point
\usepackage{bm}% bold math
\def\be{\begin{equation}}
\def\ee{\end{equation}}
\def\bea{\begin{eqnarray}}
\def\eea{\end{eqnarray}}
\def\bc{\begin{center}}
\def\ec{\end{center}}
\begin{document}
\preprint{SINP/06/2010}
\title
%{Effect of ``local'' and ``global'' trend dependencies on the stability of 
%sites in a 1D Ising chain}
{Coevolution of Glauber-like Ising dynamics on typical networks}
\author{Kamalika Basu Hajra}
\email{kamalika.hajra@saha.ac.in}
\author{Anjan Kumar Chandra}
\email{anjan.chandra@saha.ac.in}
%\author{P.K. Mohanty}
%\email{pk.mohanty@saha.ac.in}
\affiliation{%
\vskip0.2cm
$^{1}$Centre for Applied Mathematics and Computational Science and
Theoretical Condensed Matter Physics Division, Saha Institute of Nuclear
Physics, 1/AF Bidhannagar, Kolkata-700064, India.\\
}%
\date{\today}

%\def\be{\begin{equation}}
%\def\ee{\end{equation}}
%\maketitle

%\author

%{Kamalika Basu Hajra, Anjan K. Chandra and Pradip kumar Mohanty }
%\affiliation
%{
%CAMCS, TCMP Div, S. I. N. P,
%    1/AF, Bidhannagar, Kolkata 700061, India. \\
%}

\begin{abstract}

We consider coevolution of site status and link structures from two different 
initial networks: a one dimensional Ising chain and a scale free network.
The dynamics is
governed by a preassigned stability parameter $S$, and a rewiring factor
 $\phi$, that determines whether the Ising spin at the chosen site flips or 
whether
 the node gets rewired to another node in the system. This dynamics has also 
been studied with Ising spins distributed randomly among nodes which lie on a 
network with preferential attachment. We have observed the steady state average stability 
and magnetisation for both kinds of systems to have an idea about the effect
of initial network topology. Although the average stability shows almost similar behaviour, the magnetisation depends on the initial condition we start from. 
Apart from the local dynamics, the global effect on the dynamics has also been 
studied.
%dynamics has taken place for a few thousand time steps.
 These parameters show
 interesting variations for different values of $S$ and $\phi$, which
helps in determining the steady-state condition for a given substrate.
%$k_{i}$ is the degree of the $i^{th}$ node and $\tau_i$  its present age. 
%The  phase diagram in
%the ${{\alpha}-{\beta}}$ plane is obtained. The network shows scale-free
%behaviour, i.e., the degree distribution $P(k) \sim k^{-\gamma}$ with 
%$\gamma =3$
%only along a line in this plane. 
%Small world property, on the other hand, exists over a large region in the phase
%diagram. 

\end{abstract}

\maketitle
\noindent Preprint no. 

%\begin{multicols}{2}

\section {Introduction}

%     In recent times Physicists have become largely interested in the study of 
%network theory and its application as a Statistical mechanical tool to study 
%systems of diverse nature. One of the chief reasons that have interested 
%physicists in this field is the underlying behaviourial and functional
%similarity between systems which are apparently strikingly different in nature.
%This underlying similarity between systems belonging to diverse disciplines
% has brought about tremendous activity in the {\it interdisciplinary } 
%fields of science and subjects such as {\it Econophysics} and
%  {\it Sociophysics} have gained considerable importance. Just like 
%in the physics of phase transitions,
Statistical mechanics and network 
theory helps us to describe and also analyse the collective features of large 
systems such as human societies, by studying macroscopic parameters, 
 without the knowledge of the microscopic (read individual) details.
 Complex web-like structures describe a wide variety of systems of
high technological and intellectual importance. The statistical properties
of many such networks are being studied recently with much interests.
Such networks, with complex topology are common in nature and examples
include the world wide web, the Internet structure, social networks, 
communication networks, neural networks to name a few \cite{BA,DMbook,Watts_book}.
   
    In social networks,
% the interacting nodes become animate objects, 
%viz.,human beings in case of the human social networks. Evidently, 
most of the  
individuals interact with a limited number of fellow persons and this number is 
almost negligible compared to the total number of individuals comprising the 
network. In spite of this, human societies exhibit fascinating global features 
\cite{buchanan}. 
%Such features as scaling, order to disorder transition etc., 
%prompts us to use statistical mechanics as an analysing tool, as such 
%macroscopic characteristics arise due to collective phenomena occurring 
%amongst the constituting individuals, i.e., microscopic entities, in this case.
One social phenomena that is being widely explored by Physicists in recent 
times  is opinion formation or opinion dynamics.
%The goal of several social interactions is to arrive  at a common opinion
% or consensus. 
Although individual opinions in a society or group might vary, however 
after undergoing a particular dynamics, the group
 tends to present a single opinion.
% The individual opinions are most commonly 
%binary, e.g., {\it yes} or {\it no};
%, two languages in competition, 
%positive or negative reaction to a particular matter etc.
 When the individuals 
are all differing in their respective opinions, the system is
 heterogeneous 
and a Physicist would call it a `disordered system'; dynamical interaction 
would make individuals having same or similar opinion to get linked and 
those having dissimilar ideas to get detached from each other. 
It might also happen that an influential individual succeeds in altering the 
point of view of another individual in the group. When, after undergoing 
such dynamics, a consensus, or agreement is reached, the system would be 
acclaimed by a Physicist as `ordered' \cite{cast_rev}.  

One of the key facts to be kept in mind while designing  or simulating a 
social network is that in this case while the individual nodes change 
their states, the network also changes its topology due to the formation 
or severing of links between pairs of nodes. Hence not only do the nodes 
evolve, the network as a whole also evolves in time due to change 
in its link structure. Hence a correct 
representation of social dynamics should include a ``coevolution'' 
of state dynamics and network topology \cite{salvatore}. This class of 
models, where such coevolution has been studied  have been published 
profusely in past years \cite{several}. 
  
      In the present paper, we study such a system where a coevolution 
of node status and the link structure of the network takes place. Since 
binary opinion holds a major place in the opinion dynamics literature, 
we represent the agents by nodes and their opinions by spins that 
can be either plus or minus. We study different network topologies 
on which these spins are placed either randomly or in an antiferromagnetic fashion. We study a one dimensional Ising chain as well as 
a network with preferential attachment \cite{BA}. 
We study different features 
such as the average stability, magnetisation, and number of free nodes. 
%and average stability 
%to be defined shortly) with time.
 We have analysed the system from the transformation 
patterns of the above parameters.

\section {The Systems and the Dynamics}
We have studied mainly two kinds of Ising systems as our starting point: 
(i) a one dimensional Ising chain with nearest neighbour interaction and (ii) a  network grown by preferential attachment scheme, the sites of 
which are assigned by spins $\sigma = \pm 1$ with probability $1/2$. The nodes
 here are the individuals and the
spin states here represent individual opinion is  which considered to be binary (
e.g., {\em yes} or {\em no})
% placed on a one 
%dimensional chain, the spin orientation being antiferromagnetic, i.e.
%alternate sites have up or down spin. We
The dynamics follow the update rule as described below \cite{salvatore}. The 
chief parameter deciding the update rule is the 
{\it stability factor } which is defined as follows:\\
%Before stating the update rules, we define some 
%arameters of importance below:\\
{\it Stability factor} :  The stability factor $s_i$ of a particular node $i$ 
is defined as the ratio $l_i/k_i$, where $k_i$ is the number of 
links arising from the node, i.e., its degree and $l_i$ gives the number of 
neighbours having the same sign as the $i^{th}$ node. \\
Once the lattice is generated upto a desired size, say, $N$, and each
node has a particular value of spin, $l_i$, $k_i$ and $s_i$, we apply the 
update rule as follows:\\
(a) any $i^{th}$ node is selected randomly and its $s_i$ calculated.\\
(b) We denote the node to be stable if $s_i \ge S$, where $S$ is a preassigned 
value that we call ``target stability value'' and $0 \le S \le 1$. In this case 
the node does not change its sign nor does it rewire. On the other hand , 
if $s_i < S$, then a neighbour $j$ of $i$ is chosen randomly, such that 
${\sigma}_i \ne {\sigma}_j$ and \\
     (i) with a preassigned probability $\phi$ the node $i$ severes its link 
with $j$ and attaches with another node $l$ which is chosen at random from the 
rest of the network so that ${\sigma}_i = {\sigma}_l$; provided $j$ and $l$ 
were not connected previously\\
     (ii) with a probability $(1-{\phi})$ the node $i$ flips its spin.\\
% and adopts $j$'s spin state.\\
      It is worth mentioning that if during the rewiring process, any node gets 
temporarily disconnected from the network, its stability factor is assigned 
as $1$, i.e., $s_i = 1$ for such a node. This implies that a free node is
 stable and independent of the dynamics going on, until it gets connected 
during the rewiring of some other node.\\

%   The same excercise is done using a scale free network of the 
%Barabasi-Albert type,  to be described a little later.\\
  
  The chief tunable parameters here are the preassigned stability $S$ 
and the rewiring probability $\phi$. The value of $S$ determines the density 
of similar signed neighbours required for a node to be stable. The goal of 
the dynamics obviously is to reach a stable, ordered state starting from an 
initial random state.\\
%\begin{figure}[tbh]
%\noindent \includegraphics[clip,width= 6cm, angle=270]{stab_vs_phi_antev.eps}
%\noindent \includegraphics[clip,width= 6cm, angle=270]{mag_vs_phi_ppr.eps}
%\noindent \includegraphics[clip,width= 6cm, angle=270]{phi_vs_freenodes_ppr.eps}
%\end{figure}
 \section {Different Cases}
%The different cases that we study are detailed below:

\subsection {Randomly initialised one dimensional chain} 
As a starting point, the substrate chosen for the rewiring and/or spin
 flipping dynamics to take place is a  one dimensional chain of $N$
spins, where  we distribute  plus and minus (or up and down) spins randomly.
Therefore each node has an equal probability of having either $+\sigma$ or 
$-\sigma$; in our study, $|\sigma| = 1$. Each node is
connected only to its two nearest neighbours, i.e., its adjacent nodes,
 before the updating begins. 
%Due to this chosen configuration, 
% the system has a finite non zero initial magnetisation.
% the system starts off with a 
%small but finite value of $m$. 
 For each randomly chosen $i^{th}$ node, the value of the stability $s_i$ is
 determined from the values of $l_i$ and $k_i$ and the aforementioned update 
rule is applied. For this configuration, each node may initially have any
one of the following three values of $s_i$, viz., $1.0, 0.5$ and $0.0$.
Keeping  this in mind, we classify the sites in terms of their
 $s_i$ values as:\\

(i) a site whose stability is $1.0$ (connected only with other sites of same 
spin polarity) is called an
 {\em inactive site} or $i$-site. For $\phi = 0.0$, i.e., for no rewiring,
 a site whose two adjacent sites are identical is an $i$-site. 

(ii) a site whose stability is between $0.0$ and $1.0$, is called a
{\em dormant site} or $d$-site,
because these sites flip according to the value of $S$. For $\phi = 0.0$,
i.e., for no rewiring a site whose two adjacent sites are mutually opposite is
a $d$-site and the stability of such a site is always $0.5$.

(iii) a site whose stability is $0.0$ (connected only with all other sites of 
opposite spin polarity), is called an {\em active site} or $a$-site,
because these sites always flip. For $\phi = 0.0$,
i.e. for no rewiring a site whose two adjacent sites are oppositely oriented
 to the site itself is an $a$-site. 

 After allowing the system to reach the equilibrium 
configuration ($\sim 1000$ time steps), we measure the following quantities:\\
(i) The average stability per node $\langle s\rangle = {\Sigma}s_i/N$.\\
(ii) magnetisation $m = {\Sigma}_i{{\sigma}_i/N}$\\
(iii) the fraction of free nodes left $n_f$\\
%observe the variation of 
%$<s>$,$<m>$, 
%and $n_f$ with $\phi$ as earlier.  \\

\begin{figure}[tbh]

\noindent \includegraphics[clip,width= 5cm, angle=270]{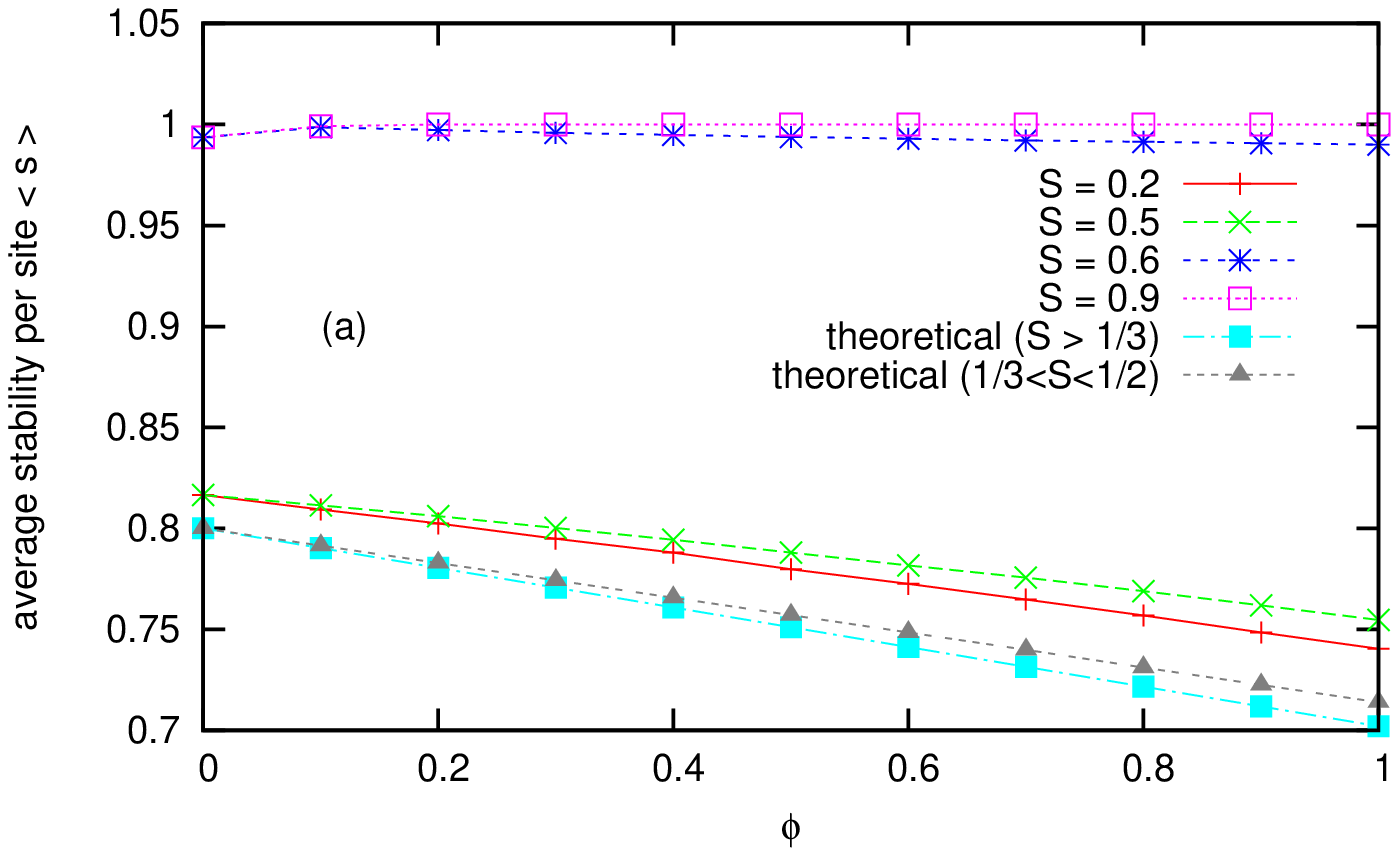}
\noindent \includegraphics[clip,width= 5cm, angle=270]{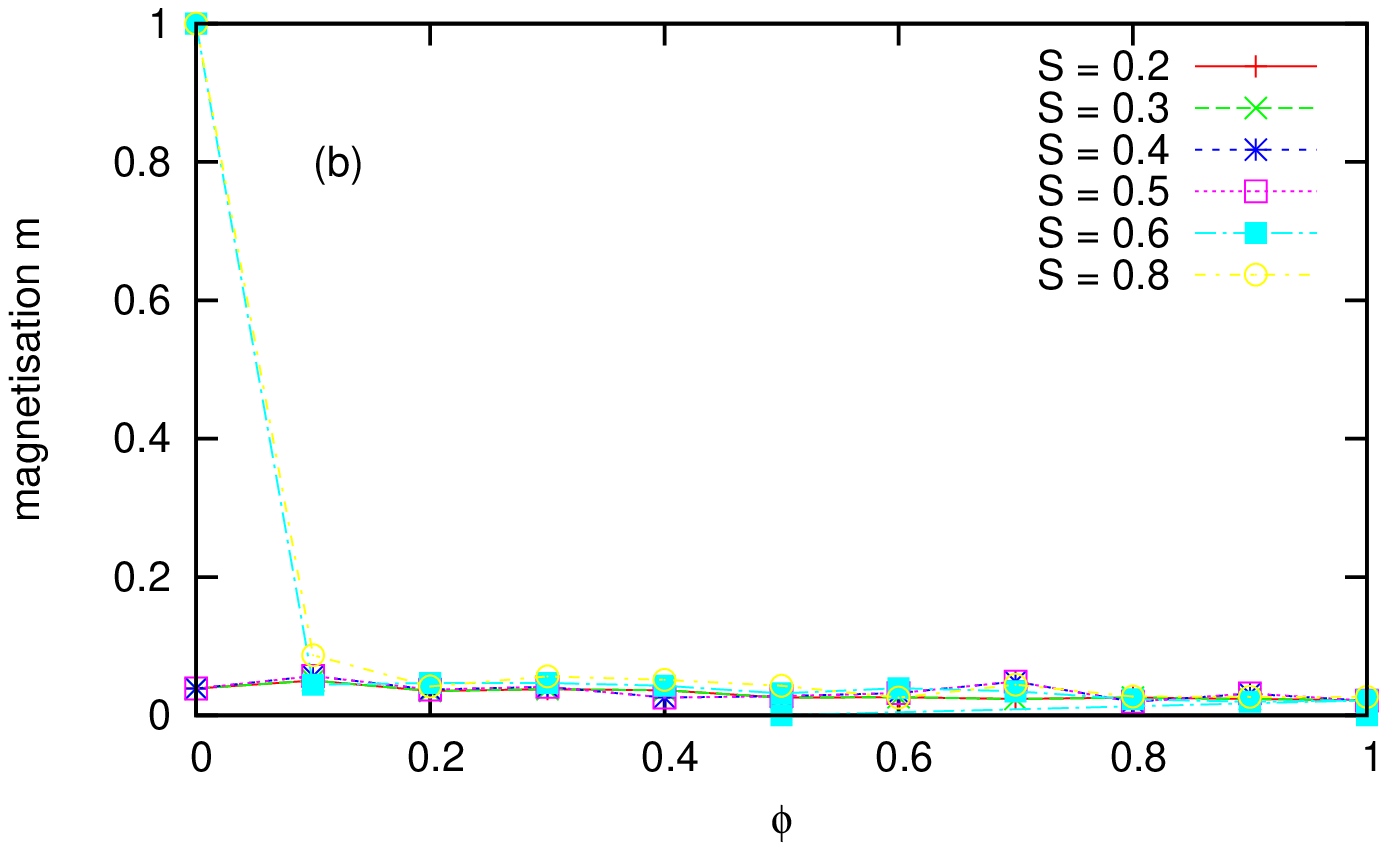}
\noindent \includegraphics[clip,width= 5cm, angle=270]{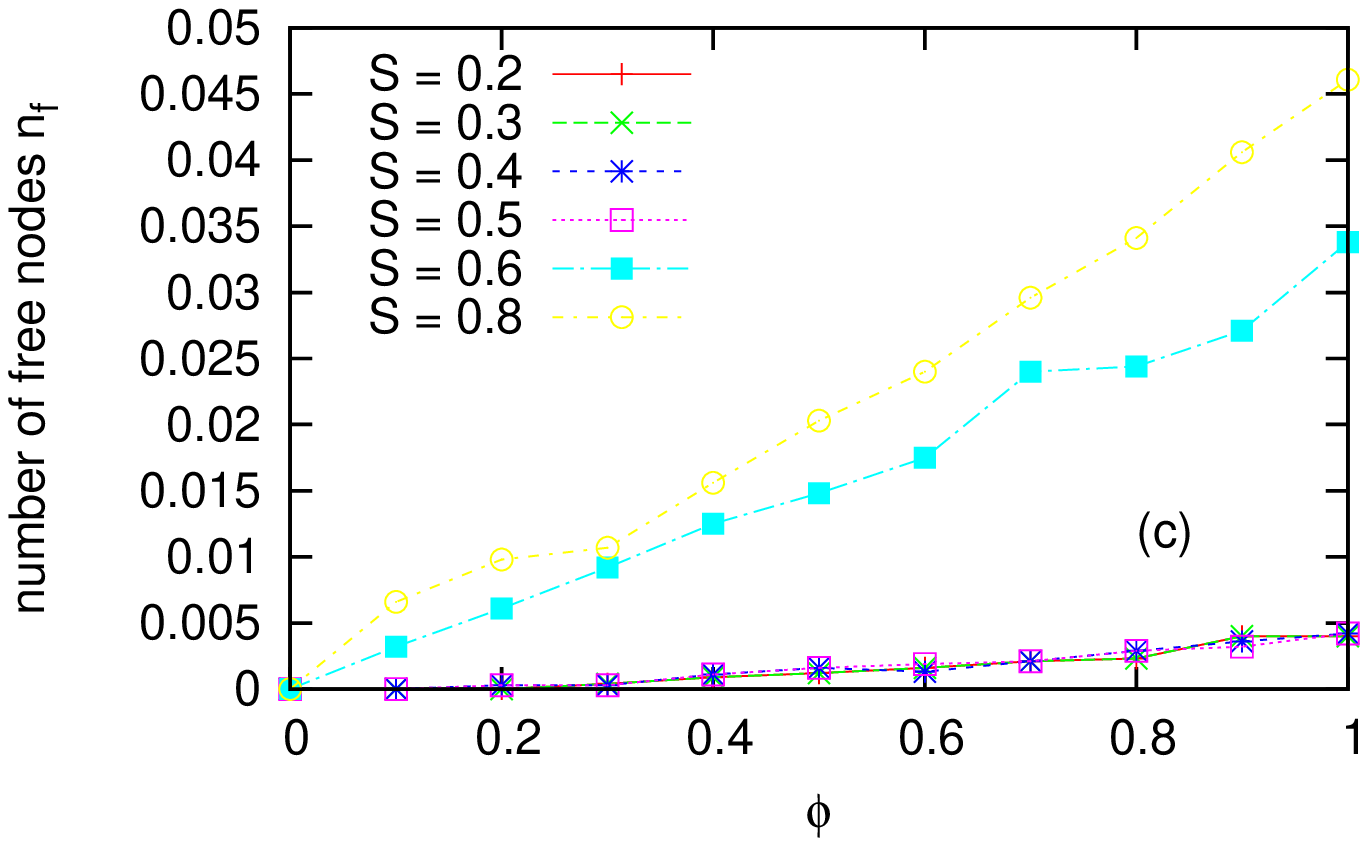}
\caption{(a) Average stability per site $\langle s \rangle$ vs $\phi$ for 
different values of $S$; the two bottom plots are the ones obtained from 
theoretical calculations for $S < 0.5$ (explanation in text) ;
(b) Magnetisation ($ m $) vs $\phi$;
 (c) Fraction of free nodes remaining in the system ($n_f$) vs $\phi$ plot
for different values of the preassigned stability factor $S$} 
\label{fig:fg1}
\end{figure}
We observe here that the average stability per node decreases with 
increasing probability of rewiring for $S \le 0.5$,  
whereas for $S > 0.5$, the value of $\langle s \rangle$ remains 
almost unaltered ($\sim 1$). For $S \le 0.5$, two branches are obtained which 
converge to the same value for $\phi = 0$ (Fig. 1a).\\

%that we start with $1/4~i$-sites, $1/4~a$-sites and $1/2~ d$-sites. For $\phi 
%\not= 0.0$ and $S < 0.5$, the change in stability will be contributed by the
%$a$-sites only (Fig. 8). When a $a$-site follows `Path I', i.e. the site
%connects with a similar site at a distance, the positive change in stability
%5occurs for three sites : the site itself ($0.5\phi$), the distant site
%($\phi[(0.25\times1/3) + (0.5\times1/6)]$) and the site with which the $a$-site
%disconnects its link ($\phi[0.50\times1/2]$). When a $a$-site follows `Path II', i.e. the site flips itself, the positive change in stability are contributed
%by three sites again : the site itself ($(1-\phi)$), the two sites with which
%the $a$-site was connected ($2(1-\phi)[0.5 + 0.5]$). So the total change in
%stability is
%\begin{equation}
%\Delta s = 3 (1 - \phi) + 0.5 \phi + \phi[(0.25\times1/3) + (0.5\times1/6)] + 0.25 \phi
%\end{equation}
\begin{figure}[h]
\noindent \includegraphics[clip,width= 8cm, angle=0]{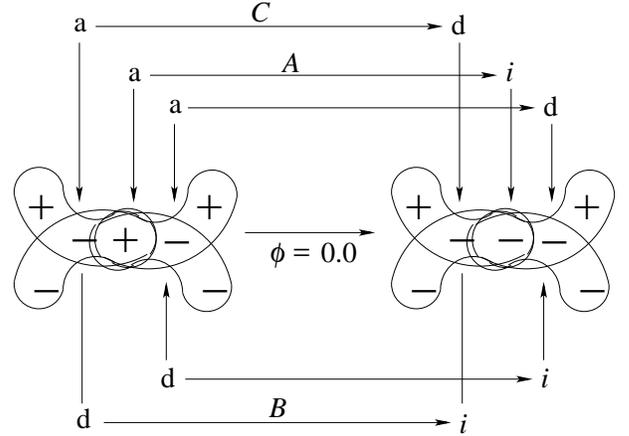}
\caption{The above figure describes how a given $a$ site undergoes 
transformation when $S < 0.5$ and $\phi = 0.0$. In the figure, an active 
site $(+)$, flips to an inactive site $(-)$ while its neighbouring sites, which 
might be active or dormant, transform to dormant or inactive sites 
respectively. Consequent change in stability corresponding to the entire 
update has been calculated to be $\Delta \langle s \rangle = 0.3$.}
%(a) average stability per site $<s>$ vs $\phi$, (b) average
%magnetisation per site $<m>$ vs $\phi$ and (c) fraction of free nodes
%$n_f$ vs $\phi$ for different values of $S$, when the initial configuration is
%a random distribution of spins on nodes which form a scale free network.}
%\noindent \includegraphics[clip,width= 5cm, angle=270]{sfn_stab_vs_phi.eps}
\end{figure}

In case of random initialisation with two links for each site, we may assume
that if there are total $N$ spins, there will be $N/4~a$ sites, $N/4~i$ sites and $N/2~d$ sites, contributing a stability of $0.50$.
Let us consider the dynamics when $\phi=0$ and $S<0.5$, i.e., when an 
$a$-site certainly converts
into an $i$-site (Fig.2. Path A). Simultaneously an adjacent $d$-site may 
convert to an 
$i$-site (Fig.2. Path B) or an $a$-site convert to a $d$-site (Fig.2. Path C)
. Initially for random configuration the
ratio of $d$-site to $a$-site is $2:1$ and conversion of each $a$-site leads
to conversion of $2$ adjacent $d$ or $a$ sites. So it is expected that
conversion of each $a$-site corresponds to the transformation of $4/3~d$ sites
to $4/3~i$ sites and $2/3~a$ sites to $2/3~d$ sites. So for a single site 
update, $5/3~a$ and $2/3~d$ sites vanish and $7/3~i$ sites appear.
The transformation equations are as follows :\\
\begin{eqnarray*}
a & \rightarrow & i ~~~~~~~(Fig 2. Path A)\\
\frac{4}{3}d  & \rightarrow & \frac{4}{3}i ~~~~~(Fig 2. Path B)\\
\frac{2}{3}a  & \rightarrow & \frac{2}{3}d ~~~~~(Fig 2. Path C)
\end{eqnarray*}
\hspace{1.5cm}\underline{~~~~~~~~~~~~~~~~~~~~~~~~~~~~}
\be
\frac{5}{3}a + \frac{2}{3}d \rightarrow  \frac{7}{3}i ~~~~~(net~ conversion) 
\ee
Initially
there are $N/4~a$ sites and the dynamics continues until they all disappear.
Instead of going into the microscopic details, we assume that the respective
sites decay at a constant rate. 
So it takes $3N/20$ steps, and during this time $3N/20 \times 2/3 = N/10~d$ 
sites vanishes. So finally, $2N/5~d$ sites are left and the remaining are
$3N/5~i$ sites leading to a stability of $0.8$.
\begin{figure}[h]
\noindent \includegraphics[clip,width= 8cm, angle=0]{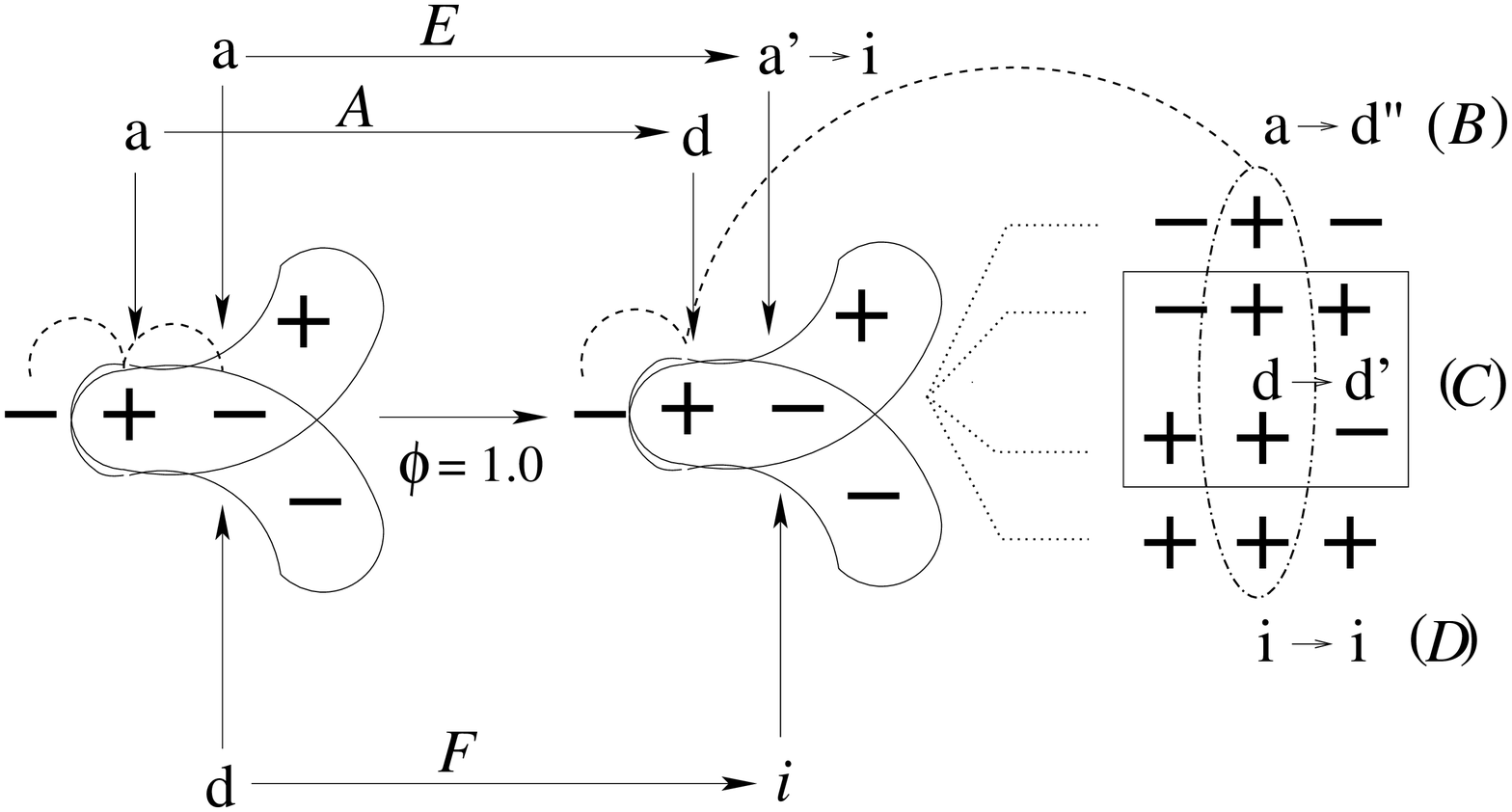}
\caption{When $\phi = 1.0$, there is no spin flip, but an $a$ site would sevre its link from one of its neighbours and link to a distant site, which might be $a$, $d$, or $i$. The above figure shows such a transformation and depicts 
how the configuration of the distant site might change due to addition of a 
like sign-ed node and the four possibilities are shown. The original
 neighbour, from whom the link has been severed will also change to an
 $a$ or an $i$ site. In this case, the theoretically calculated value of 
$\Delta \langle s \rangle = 0.202$ or $0.214$ (details in text)}
%an $a$-site "A" severes its link with its neighbour and joins a distant
 %site "B" which might be $a$, $i$ or $d$ site. The corresponding transformations  of all the participating sites is shown. e.g., if "B" is inactive, it remains so, while if it is active, it becomes dormant and its $s_i$ changes from $0$ to 
%$1/3$.} 
%magnetisation per site $<m>$ vs $\phi$ and (c) fraction of free nodes
%$n_f$ vs $\phi$ for different values of $S$, when the initial configuration is
%a random distribution of spins on nodes which form a scale free network.}
%\noindent \includegraphics[clip,width= 5cm, angle=270]{sfn_stab_vs_phi.eps}
\end{figure}

Now let us concentrate on the dynamics followed by the system for $\phi=1.0$
and $S<0.5$, i.e., when stability increases only through the process of 
rewiring. In this
case also the $a$-sites take the major role, while the other sites are 
affected
indirectly. A distinct property that largely discriminates this dynamics from
the previously described one, is that during a single update only one of the 
adjacent sites get
affected due to the conversion of the main $a$-site and the other one remains
totally undisturbed. Instead of that a site at an arbitrary distance from the
principal site with which it gets newly connected may give rise to various
configurations depending upon the initial state it starts off. Let us
go through all the possible changes one by one. The compulsory change is
that of an $a$-site converting to a $d$-site (Fig.3 Path A) by getting 
disconnected from any of
the adjacent oppositely oriented spin and rewiring to a distant one of the
same orientation. The distant spin may be an $a$-site with probability $1/4$
and stability $0.0$, $d$-site with probability $1/2$, stability $0.5$ or an
$i$-site with probability $1/4$ and stability $1.0$ (shown in Fig.3 by an 
elliptical boundary). Now due to the rewiring
it changes respectively to a dormant site (say $d^{\prime\prime}$) with
stability $1/3$ (Fig.3 Path B), a dormant site (say $d^{\prime}$) with 
stability $2/3$ (Fig.3 Path C) or
an $i$ site (Fig.3 Path D) with no change in stability. On the other hand, 
the node from
which a link is disconnected may be either an $a$-site with probability $1/2$
or a $d$-site with probability $1/2$. The disconnected $a$-site transforms to 
an active site 
although with only one link (say $a^{\prime}$)  
(Fig.3 Path E) or a $d$-site to an $i$ site with
stability $1.0$ (Fig.3 Path F). However the $a^{\prime}$ hardly remains 
stable and quickly
transforms to an $i$ site (Fig.3 Path E).
\begin{eqnarray*}
a & \rightarrow & d ~~~~~~~~~~~~~~\mbox{(Fig 3. Path A)}\\
\frac{1}{4}a  & \rightarrow & \frac{1}{4}d^{\prime\prime} ~~~~~~~~~~~\mbox{(Fig 3. Path B)}\\
\frac{1}{2}d  & \rightarrow & \frac{1}{2}d^{\prime} ~~~~~~~~~~~~\mbox{(Fig 3. Path C)}\\
\frac{1}{2}a  & \rightarrow & \frac{1}{2}a^{\prime} \rightarrow \frac{1}{2}i ~~~~~\mbox{(Fig 3. Path D)}\\
\frac{1}{2}d  & \rightarrow & \frac{1}{2}i ~~~~~~~~~~~~~\mbox{(Fig 3. Path E)}
\end{eqnarray*}
\hspace{1.0cm}\underline{~~~~~~~~~~~~~~~~~~~~~~~~~~~~~~~~~}
\be
\frac{7}{4}a \rightarrow  \frac{1}{4}d^{\prime\prime} + \frac{1}{2}d^{\prime}
+ i ~~~~~(net~ conversion)
\ee
Hence the net transformation leads to the reduction of $7/4~a$ sites at each
step and appearance of $1/4~d^{\prime\prime}$, $1/2~d^{\prime}$ and
and an $i$- site. Again we can calculate approximately (ignoring the rigorous
dynamics that really takes place) the increase in stability due to this
transformation. It takes $N/7$ steps for all the $N/4~a$ sites to vanish. 
During this time, the new sites that appear are
$N/28~d^{\prime\prime}$, $N/14~d^{\prime}$ and $N/7~i$ sites. So the increase
in stability is $(1/28\times 1/3) + (1/14\times 2/3) + (1/7\times 1) = 0.202$. As we begin with a stability of $0.5$, the final value we approach is $0.702$ which is again $0.02$ lower than that obtained from simulation.
Now we can estimate the $\langle s \rangle$ values for all other values
of $\phi$ from these two limiting values of stability enhancement.
At $\phi = 0.0, \Delta \langle s \rangle = 0.3$ and at $\phi = 1.0$,
$\Delta \langle s \rangle = 0.202$. For any value of $\phi, \Delta \langle s 
\rangle = 0.3(1 - \phi) + 0.202\phi$.
The two branches for $S < 0.5$ can also be explained from the instability
of the $d^{\prime\prime}$ sites when $S > 1/3$. In that case for $\phi = 1.0$ the
$d^{\prime\prime}$ sites get immediately transformed to $d^{\prime}$ sites
of stability $2/3$ and thus $\Delta \langle s \rangle = (1/28\times 2/3) + 
(1/14\times 2/3) + (1/7\times 1) = 0.214$ and thus for any arbitrary $\phi$,
$\Delta \langle s \rangle = 0.3(1 - \phi) + 0.214\phi$. These two analytical
lines have been shown in Fig.1(a) and are very close to those obtained from
simulation.

The magnetisation shows a considerably high value $\sim 1$ for values 
of $S > 0.5$, only at $\phi = 0.0$ (Fig. 1b). However the value of 
magnetisation is very low otherwise.

% which was not observed in our earliesubstrate.
 The reason for the high value can be understood with a 
little insight. For $\phi=0$, no rewiring occurs and the dynamics proceed
 only through flipping of spins.
%If we consider each spin with its two nearest neighbours, three types of
%sites can be found :

\begin{figure}[h]
\begin{center}
\noindent \includegraphics[clip,width= 5cm, angle = 270]{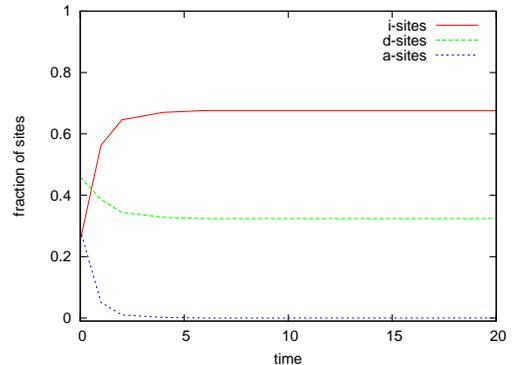}
\caption{The plot of time vs fraction of different sites, (i.e., i, d and a sites) for $S = 0.3, \phi = 0.0$}
\label{fig:fg2}
\end{center}
\end{figure}

\begin{figure}[h]
\begin{center}
\noindent \includegraphics[clip,width= 5cm, angle = 270]{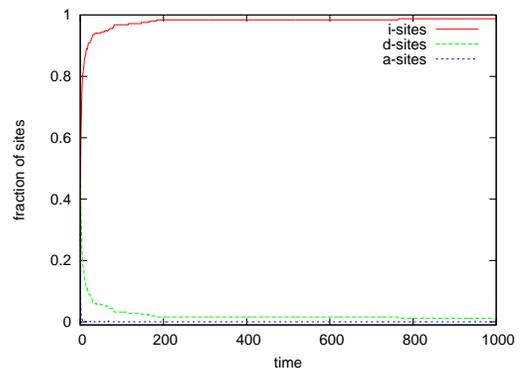}
\caption{The plot of $time - fraction$ for different values of $S=0.6, \phi = 0.0$}
\label{fig:fg3}
\end{center}
\end{figure}

%(i) a site with identical nearest neighbours ({\em i.e.}
%$+++$ or $---$), which we say {\em inactive sites} or $i$-sites, because these
%sites remain unflipped for ever.

%(ii) a site with two nearest neighbours oriented along mutually opposite
%directions ({\em i.e.} $++-$ or $--+$ or $-++$ or $+--$), which we designate
%as {\em dormant sites} or $d$-sites, because these sites may flip (depending
%upon the value of the assigned $S$).

%(iii) a site with two nearest neighbours oriented opposite to that of the spin
%of that site ({\em i.e.} $-+-$ or $+-+$), which we will say {\em active sites}
%or $a$-sites, because these sites always flip and convert to an $i$-site
%(independent of the value of the $S$).

According to the definition of the $i, d$ and $a$ sites, the difference in 
dynamics at $\phi = 0.0$ is due to the activity of the $d$-sites.
When $S<0.5$, most of the $d$-sites (whose stability factor is $0.5$) remain 
dormant
forever (Fig.~\ref{fig:fg2}). From our previous calculation, it can be approximated that $4/5$ of the $d$-sites persist. Thus once all the $a$-sites are 
updated, the dynamics stop (even
though a large fraction of $d$-sites remain intact as we begin with a random
configuration). However some adjacent sites transform due to indirect effect.
For example, during the update process, while an $a$-site 
converts to an $i$-site, an adjacent $d$-site transforms to an
$i$-site. But as a whole, due to the presence of a large fraction of $d$-sites
(in the equilibrium configuration), the magnetisation is very small.

The situation drastically changes for $S>0.5$. For $S>0.5$, the $d$-sites always
flip and converts to another $d$-site (to fulfil the stability criterion)
with a different configuration ({\em i.e.}
 $++-$ becomes $+--$ or $--+$ becomes $-++$). So the domain walls perform a
random walk until they annihilate each other and all the sites become 
inactive asymptotically (Fig.~\ref{fig:fg3}).
Obviously the $a$-sites also transform to $i$-sites.
So as $S$ exceeds the value of $0.5$, all the spins become either up or down
and thus the magnetisation reaches the value $1.0$.\\
The plot of the fraction of free nodes left in the system, $n_f$, vs $\phi$ 
is shown in Fig. 1c. The value of $n_f$ increases with increasing probability 
of rewiring, but the nature of increase shows a marked difference for values of 
$S \le 0.5$ and $S > 0.5$.\\ 

\begin{figure}[tbh]
\noindent \includegraphics[clip,width= 5cm, angle = 270]{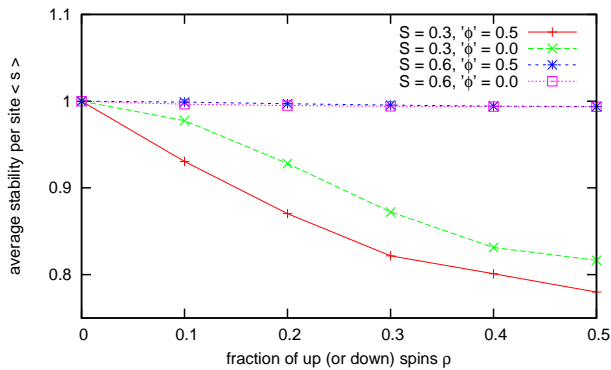}
\caption{Average stability per site versus fraction of up (or down) spins
$b$ at $\phi = 0.0$ and $\phi = 0.5$ for $S = 0.3$ and $S = 0.6$}
\label{fig6}
%\end{center}
\end{figure}

Another interesting variation that we observed in this case was the 
variation of the value of $\langle s\rangle$ with the fraction of up spins 
($\rho$). 
$\rho = 0$ means all the spins are down and 
$\rho = 1/2$ means equal number of up and down spins. This 
variation is measured for different values of $S$ and $\phi = 0$, i.e., 
zero probability of rewiring (Fig.~\ref{fig6}). 
It is observed that for values of $S \leq 0.5$ the average stability 
per node $\langle s\rangle$ decreases with $\rho$ and reaches a minimum when 
$\rho \approx 0.5$.
%, and then again increases with further increase in $\rho$. 
For values of $S > 0.5$ however, $\langle s\rangle$ remains almost constant 
($\sim 1$) with increasing $\rho$.\\
 
\subsection{One dimensional chain with antiferromagnetic initialisation}

\begin{figure}[tbh]
\noindent \includegraphics[clip,width= 5cm, angle=270]{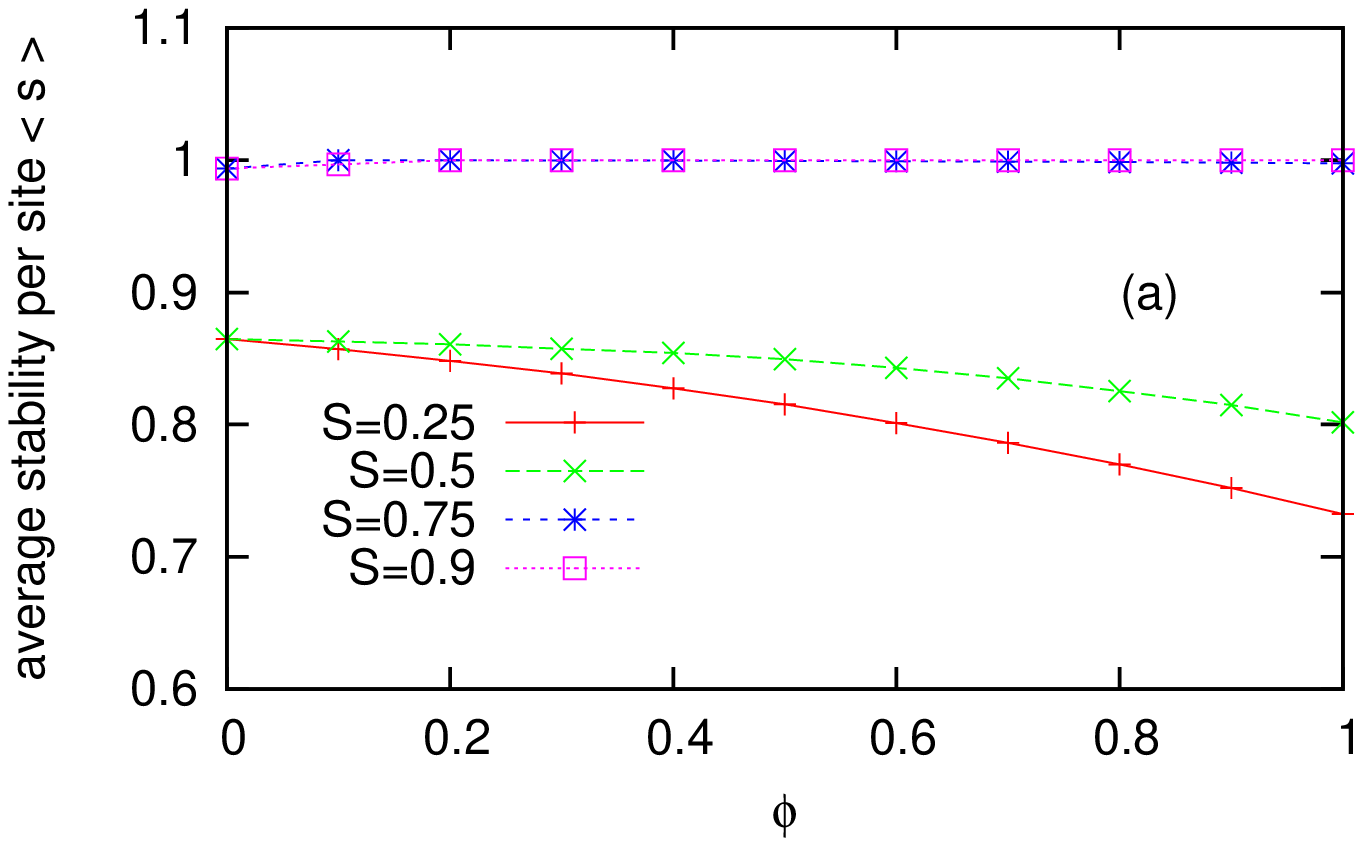}
\noindent \includegraphics[clip,width= 5cm, angle=270]{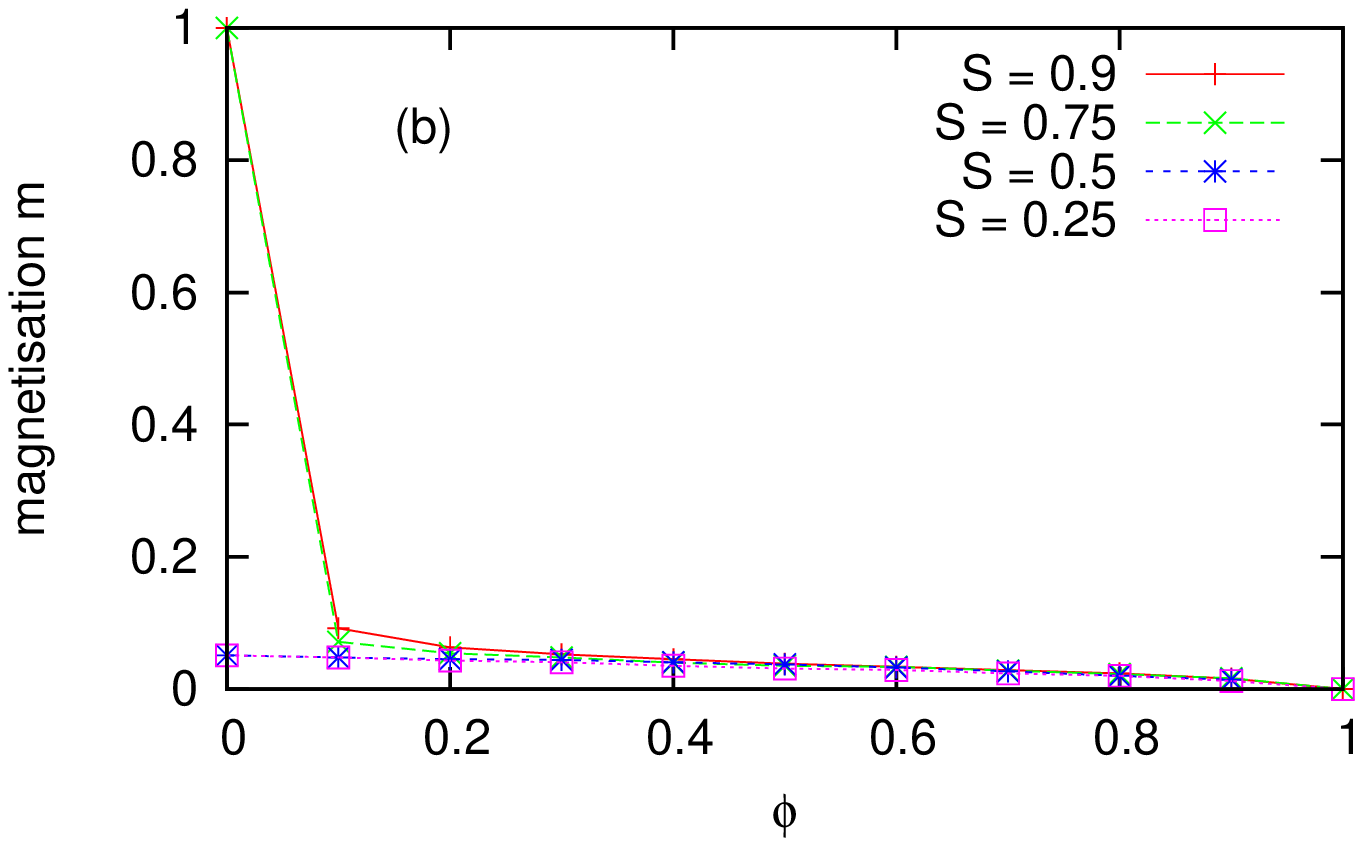}
%\noindent \includegraphics[clip,width= 5cm, angle=270]{antev_freenodes_vs_phi.eps}
\caption{(a) average stability per site $\langle s\rangle$ vs $\phi$, (b) average 
magnetisation per site $m$ vs $\phi$} 
%and (c) fraction of free nodes $n_f$ 
%vs $\phi$ for different values of $S$, when the initial configuration is 
%n{figure}[h]
%\noindent \includegraphics[clip,width= 8cm, angle=0]{transform.eps}
%\caption{(a) average stability per site $<s>$ vs $\phi$, (b) average
%magnetisation per site $<m>$ vs $\phi$ and (c) fraction of free nodes
%$n_f$ vs $\phi$ for different values of $S$, when the initial configuration is
%a random distribution of spins on nodes which form a scale free network.}
%\noindent \includegraphics[clip,width= 5cm, angle=270]{sfn_stab_vs_phi.eps}
%\end{figure}antiferromagnetic}
\end{figure}
This is a special case of the previously described lattice
 where the initial configuration is a 
one dimensional chain of nodes with alternate sites having spin $+{\sigma}$ 
and $-{\sigma}$. 
As mentioned earlier, $|{\sigma}| = 1$ and initially each node is connected 
with only nearest neighbours. We have separately studied this special case
for two reasons : (i) to see how much the results vary with the previous one if
we start off with a periodic array of spins and (ii) some results have already
been derived exactly for the limiting case ($\phi = 0.0$).
Evidently, $s_i = 0$ for all nodes as in this case we start 
off with $l_i =0$ and $k_i =2$ for all $i$. Therefore no matter how small a 
value of $S$ we assign, a dynamics will take place to approach stability. 
Once the system reaches the equilibrium configuration 
 we measure $\langle s\rangle$, $m$ and $n_f$, as defined earlier
%the following quantities:\\
%(i) magnetisation $m = {\Sigma}_i{{\sigma}_i/N}$\\
%(ii) the number of free nodes left $n_f$\\
%(i) The average stability per node $<s> = {\Sigma}S_i/N$.\\
%(ii) magnetisation $m = {\Sigma}_i{{\sigma}_i/N}$\\
%(iii) the fraction of free nodes left $n_f$\\

We show in Fig. 7(a) the $\langle s\rangle$ vs $\phi$ plots for different 
values of $S$. It is observed
 that we get three distinct branches for the various values of the preassigned 
stability factor $S$. For values of $S \le 0.5$, we get two branches, and 
 $\langle s\rangle$ decreases with increasing value 
of $\phi$, however the two branches converge to the same value of $\langle s\rangle$ at $\phi = 0$,
 namely  ${\langle s\rangle}_{(\phi=0)} \sim 0.86$. When we start with an antiferromagnetic 
configuration of spins, all sites are $a$-sites, so that when $S < 0.5$, 
random updating leads to a final configuration that consists of domains of 
size greater than or equal to three. Updating of each  $a$-site gives rise 
to a domain of odd number of sites and this process continues until all the
 $a$-sites vanish. It has been  analytically proved that, in the steady 
state, the fraction of domain walls approaches a value of $1/{e}^2$ 
\cite{flory,JCP}. So fraction of $d$-sites reach the value 
$2/{e}^2 = 0.2706$ and the remaining $0.7294$ are $i$-sites. Since
 stability for a $d$-site is $0.5$ and that for an $i$-site is $1.0$, 
 the stability factor approaches a value 
$0.5 \times 0.2706 + 1.0 \times 0.7294 = 0.8647$. On the other hand, when we start with a 
random initial configuration of spins, with $S < 0.5$, the fraction of 
$a$-sites starts from $0.5$ and saturates at a value which is slightly 
higher ($0.3243$) than that obtained for antiferro initialisation. Consequently 
the fraction of $i$-sites decreases and thus the overall stability factor 
becomes lower ($0.83785$) as seen in Fig. 1(a).
% [explanation to be given with ref from JCP and 
\\
 The average magnetisation per site as we can see, is $1.0$, only for $\phi = 0.0$ and $S>0.5$. Otherwise for any value of $\phi$ and $S$, it is very low. So
overall it is qualitatively same as for random initialisation.

% then at $\rho = 0$, all 
%spins are either up or down and at $\rho = 1 $, vice versa. This 
%variation is measured at different values of $S$ for $\phi = 0$, i.e., 
%zero probability of rewiring. 
%It is observed that for values of $S \leq 0.5$ the average stability 
%per node $s_i$ decreases with $\rho$, reaches a minimum when 
%$\rho \approx 0.5$, and then again increases with further increase in $\rho$. 
%For values of $S \geq 0.6$ however, $<s>$ remains almost constant with
% increasing $\rho$. 
%[explain***]\\

%\begin{figure}[tbh]
%\noindent \includegraphics[clip,width= 5cm, angle = 270]{stability_vs_ratio_rand.eps}
%\caption{The plot of $time - fraction$ for different values of $s=0.6$}
%\label{fig:fg3}
%\end{center}
%\end{figure}

%\pagebreak
\subsection{Effect of global magnetisation}

In this subsection we intend to study the effect of global magnetisation on 
the system. Till now the 
spin flip, or rewiring with a distant node with same spin was dependent on the 
value of $s_i$ and assigned $\phi$. However
 we observe significant changes if instead of local dependence, we introduce a 
global effect in the dynamics. Now a selected site will flip with probability
$1-\phi$, only if its spin
does not match with the sign of the global magnetisation, i.e. the 
magnetisation of the system. In other words, it
 may so happen that even though a situation arises when flipping of the spin
increases stability, the global magnetisation prohibits the system to gain that
enhanced stability. Since global trends often appear as strong driving factors in 
societies, introduction of this global dependence makes our study more 
realistic. We observe the variation of the  
parameters $\langle s\rangle$ (Fig. 8a) and $m$ (Fig. 8b) with the rewiring 
probability $\phi$ for 
different values of $S$. The most striking observation in this case is that 
both the average stability and magnetisation per site retains a considerably 
high value for a wide range of the rewiring parameter $\phi$ for $S > 0.5$ .  
So it can be inferred that following the global trend not only retains the
high 
stability value but also brings about homogeneity to the system. Nevertheless 
the striking difference in the system behaviour for values of $S \le 0.5$ and 
$S > 0.5$ is once again apparent from all the plots. \\

\begin{figure}[tbh]
\noindent \includegraphics[clip,width= 5cm, angle=270]{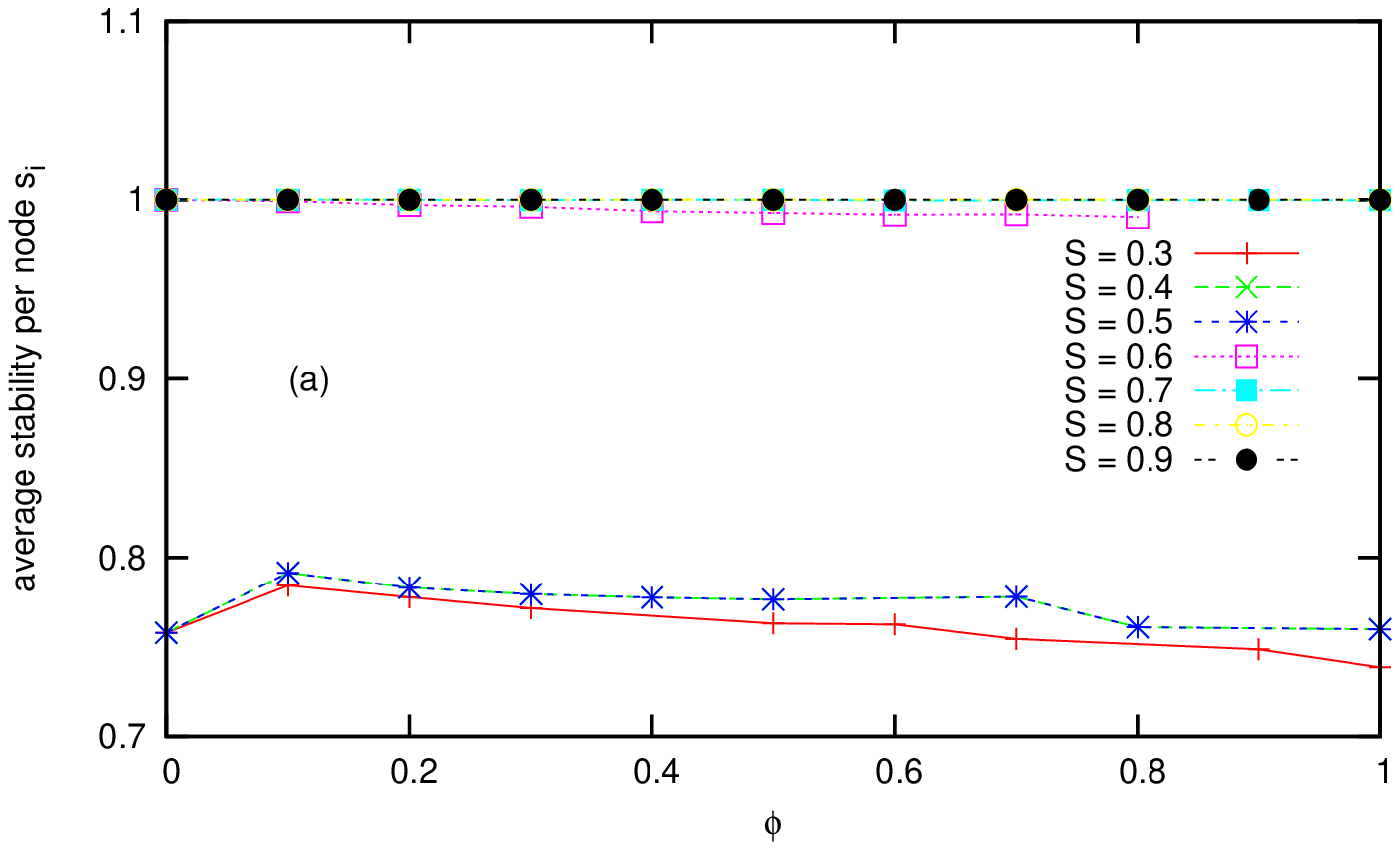}
\noindent \includegraphics[clip,width= 5cm, angle=270]{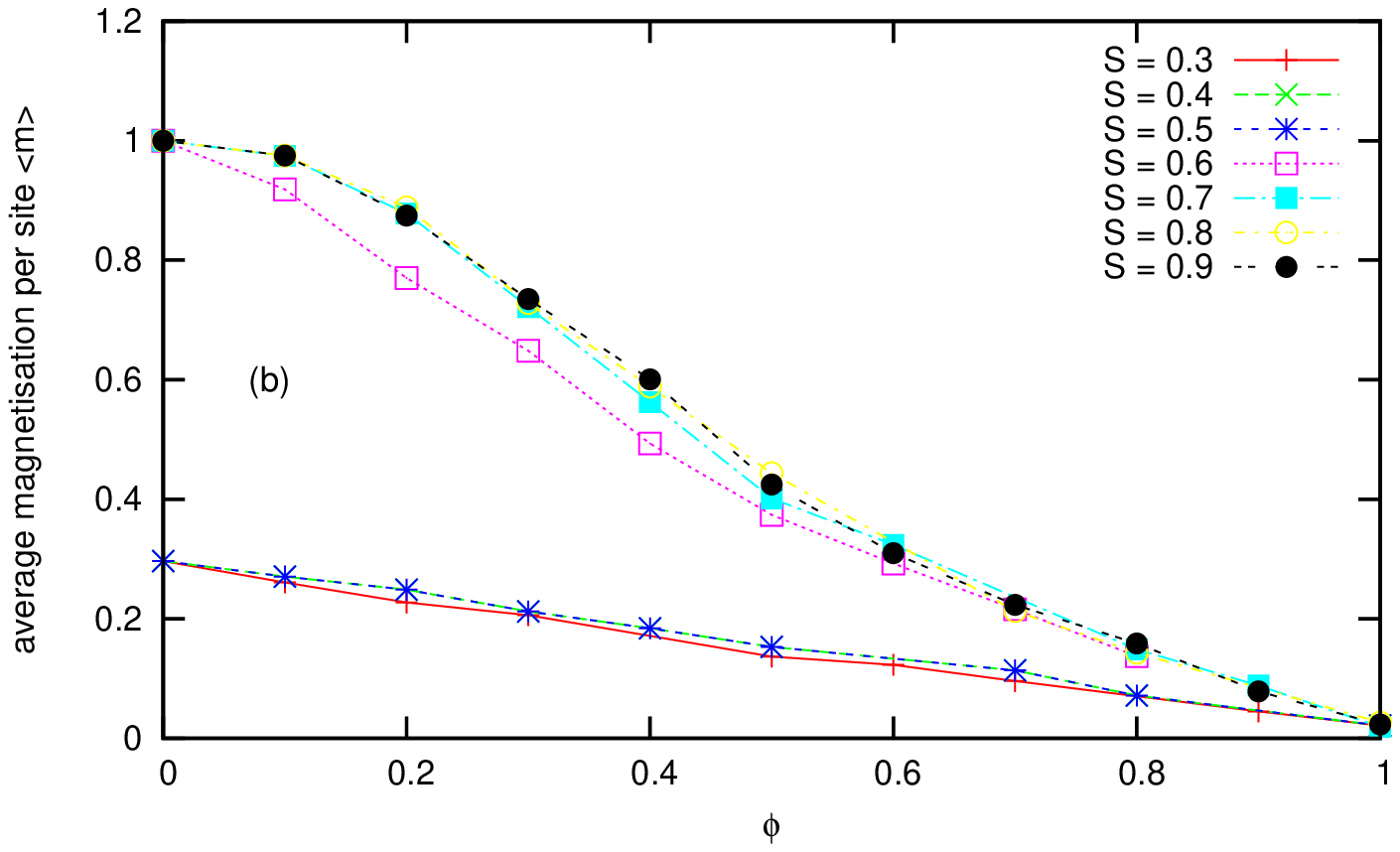}
%\noindent \includegraphics[clip,width= 5cm, angle=270]{glob_freenodes.eps}
%\noindent \includegraphics[clip,width= 5cm, angle=270]{glob_stab_vs_phi.eps}
\caption{(a) Average stability per node $\langle s\rangle$ vs $\phi$, (b) Average 
magnetisation per site $m$ vs $\phi$} 
%and (c) fraction of free nodes $n_f$ vs 
%$\phi$ for different values of $S$ when the spin flip dynamics follows the 
%global magnetisation}
\end{figure}

\subsection{Network with preferential attachment}
 It is a well known fact that a large number of real systems show 
the topology of a {\it Scale free network} \cite{BA}. In a nutshell, 
a scale free network is one in which the connection probability of a new node 
to an existing node is proportional to the degree 
(or number of links/neighbours) of the existing node,i.e.,
\begin{equation}
{\Pi}_i \sim k_i
\end{equation} 
at a given timestep. Here, 
the attachment is preferential instead of being random. For such networks, the 
degree distribution follows a power law, viz. $P(k)\sim k^{\gamma}$
and such networks are characterised 
by the existence of {\it hubs}, i.e., few nodes with very high concentration of 
links. Scale free networks form an extremely important genre of study for 
network theorists as several real world networks belong to this class. Keeping 
these in mind we next use a fully evolved scale free network as the substrate 
on which we place up or down spins on the nodes and carry out the dynamics 
mentioned earlier.\\

\begin{figure}[tbh]
\noindent \includegraphics[clip,width= 5cm, angle=270]{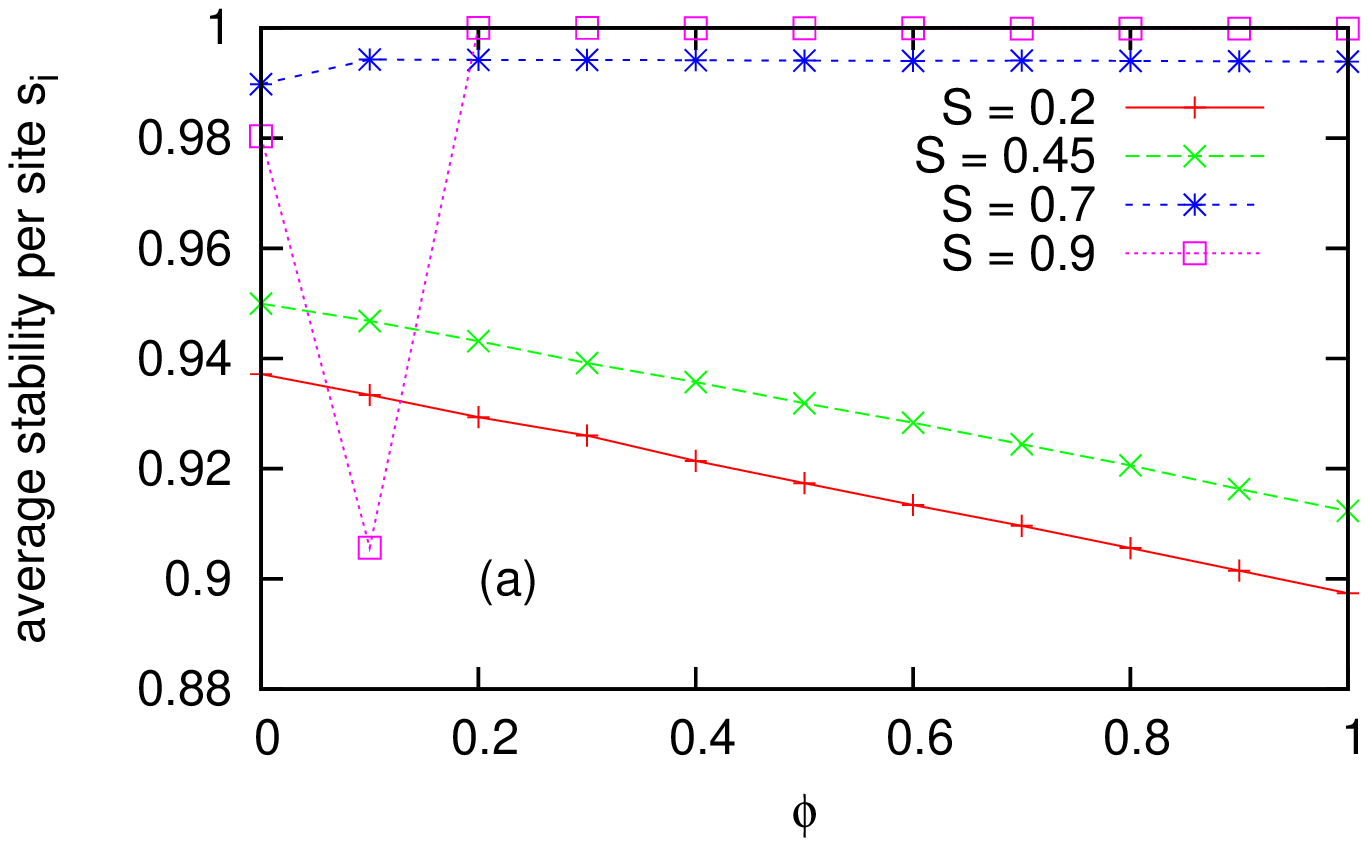}
\noindent \includegraphics[clip,width= 5cm, angle=270]{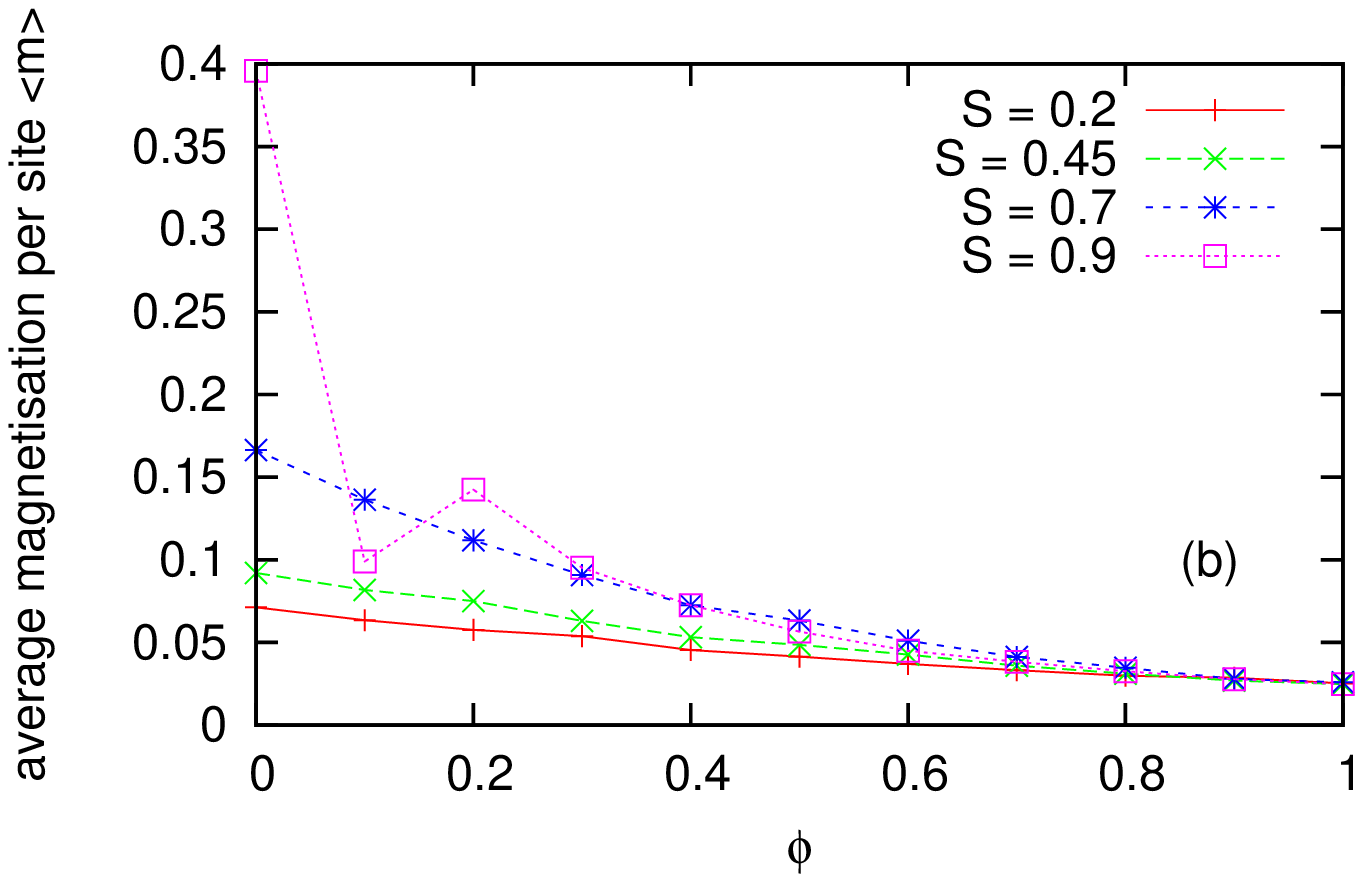}
%\noindent \includegraphics[clip,width= 5cm, angle=270]{sfn_freenodes.eps}
\caption{(a) average stability per site $\langle s\rangle$ vs $\phi$, (b) average 
magnetisation per site $m$ vs $\phi$} 
%and (c) fraction of free nodes 
%$n_f$ vs $\phi$ for different values of $S$, when the initial configuration is 
%a random distribution of spins on nodes which form a scale free network.}
%\noindent \includegraphics[clip,width= 5cm, angle=270]{sfn_stab_vs_phi.eps}
\end{figure}

A very important modification to the Barab\'asi-Albert type network is the one where the attachment probability has a nonlinear dependence on the degree 
\cite{KRL}, i.e.,
 \begin{equation}
{\Pi}_i \sim {k_i}^{\beta}
\end{equation}
 In this case, it may be shown that the network is scale free, i.e., 
the degree distribution is a power law only for linear dependence, when 
$\beta = 1.0$. Such 
nonlinear modifications have been studied in details in \cite{onody, KBH}. We 
made an investigation to find out any change in the magnetisation  
if the nonlinear degree dependence is introduced, when the system behaves 
as a small world instead of scale free. We found that indeed there is a drastic 
change in the magnetisation, which now showed considerably high values for 
$\beta > 1.0$ with $S = 0.9$ and $\phi$ varying from $0$ to $1.0$ (Fig.10). 
This conclusively shows that the system parameters are dependent on the 
initial configuration of the substrate.
\begin{figure}[tbh]
\noindent \includegraphics[clip,width= 5cm, angle=270]{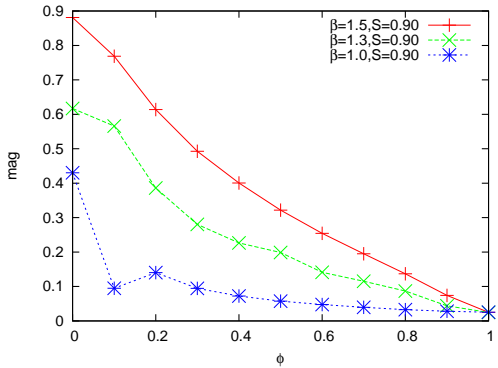}
\caption{Plot of magnetisation vs ${\phi}$ for $S=0.9$ for three different 
values of network parameter $\beta$. Interestingly, for non linear dependence 
of the connection probability on degree of a node, considerably high values 
of magnetisation is obtained even for $\phi > 0$}
\end{figure}
% If we start with a random distribution of spins among the nodes, with 
%approximately  half of the spins being up and the other half down, then
% initially the system magnetisation will be almost zero.
% We measure once again the 
%average stability per site ,i.e., $\langle s\rangle$, the number of disconnected components 
%left in the system after the dynamics and the average magnetisation per site.
%The plots are shown in the Fig.s 7(a) and (b).\\

\section{Discussions}

We have addressed here  a simple model undergoing coevolution of node status 
and link structure. We have used the same update rule as \cite{salvatore} on 
different kinds of initial substrates. Spins randomly placed on nodes on a 
one dimensional lattice has been studied in some details where we have 
not only presented numerical results but have also tried to put forward a 
theoretical explanation of the same. Simulations have also been made for 
the same lattice structure but when  spins are placed in an 
antiferromagnetic fashion initially. The target stability can be thought of 
as a measure of the number of ``like-minded" neighbours a particular 
agent should have in order to be called stable. Obviously, when the target 
stability is small, only the active sites ($a$ sites)  undergo dynamics and 
overall stability of the system shows a decrease. When the spin dynamics is 
considered and changes in $s$ are calculated from corresponding rate equations,
 the theoretical and numerical results  match considerably well albeit with 
slight difference in values.
For the case where we have considered a network grown following the preferential attachment scheme, it is observed that although for scale free behaviour of the network, the variation of magnetisation with $\phi$ does not show any 
significantly different behaviour, however as soon as we enter the non linear 
region, where according to \cite{KRL,KBH}, scale free nature disappears and 
small world behaviour predominates and a `` gel '' formation takes place, we find the magnetisation to reach considerably high values even when $\phi > 0.0$. This conclusively shows that the system parameters depend on initial configuration of the agents. \\

\begin{acknowledgements}
The authors acknowledge many fruitful discussions and suggestions of Prof. P.K. Mohanty and Soumyajyoti Biswas. The computational facilities of CAMCS of 
SINP were used in producing the numerical results.
\end{acknowledgements}

\pagebreak

% Acknowledgments: We thank S. S. Manna for useful comments. KBH is grateful to CSIR (India) F.NO.9/28(609)/2003-EMR-I for financial support.
%PS acknowledges DST grant no.  SP/S2/M-11/99.

%Email: kamalikabasu2000@yahoo.com,  parongama@vsnl.net

\end{document}